# Machine learning meets Singular Optics: Speckle-based Structured light demultiplexing

Venugopal Raskatla[1,2*], Purnesh Singh Badavath[1], and Vijay Kumar[1,†]

[1]Department of Physics, National Institute of Technology, Warangal, Telangana, 506004, India
[2]Current address: Optoelectronics Research Centre and Centre for Photonic Metamaterials, University of Southampton, Southampton, SO17 1BJ, UK

Email: [*]venuraskatla@gmail.com, [†]vijaykumarsop@gmail.com

**Abstract:**

In this paper, the advancements in structured light beams recognition using speckle-based convolutional neural networks (CNNs) have been presented. Speckle fields, generated by the interference of multiple wavefronts diffracted and scattered through a diffuser, project a random distribution. The generated random distribution of phase and intensity correlates to the structured light beam of the corresponding speckle field. This unique distribution of phase and intensity offers an additional dimension for recognizing the encoded information in structured light. The CNNs are well-suited for harnessing this unique ability to recognize the speckle field by learning hidden patterns within data. One notable advantage of speckle-based recognition is their ability to identify structured light beams from a small portion of the speckle field, even in high-noise environments. The diffractive nature of the speckle field enables off-axis recognition, showcasing its capability in information broadcasting employing structured light beams. This is a significant departure from direct-mode detection-based models to alignment-free speckle-based detection models, which are no longer constrained by the directionality of laser beams.

**Keywords:** Orbital Angular Momentum Beams, Structured Light, Speckle, Deep Learning, Convolutional Neural Network

## 1. Introduction:

Structure light - a spatially tailored light, paves many applications ranging from optical trapping to communication[1,2]. Due to their multiple degrees of freedom (wavelength, polarization, frequency, amplitude, etc) to tailor, they have become a core interest to many researchers and engineers[3]. Laguerre–Gaussian (LG) beams – a special subset of structured light carrying orbital angular momentum (OAM), provide a theoretically infinite discrete basis set due to their quantized helical phase. The LG beams and other classes of structured light such as Hermite-Gaussian (HG), and their superposition modes can be generated by illuminating the Gaussian laser beam on SLMs or DMDs displaying their fork holograms. On the other hand, they can be classified by looking at their interference and diffraction patterns. The accuracy of detection using these traditional methods is limited by alignment issues, noise, and beam wandering. Almost a decade ago, artificial intelligence-based structured light recognition models were proposed and used in optical communication links[4]. These models uplifted the constraints of traditional detection methods by reducing the complexity, boosting the speed of detection, and automating the process. Later, several machine learning and deep learning models have been proposed to make the recognition accuracy robust to noise and other perturbations in the beam profile. But still, these machine learning models required the whole beam profile to make accurate classification.

In this paper, we are presenting speckle-based models employing convolutional neural network (CNN) for the recognition of structured light modes[5–8]. These speckle-learned models overcome the limitation of previous machine-learning techniques by further lifting the alignment constraint. Even though speckle fields visually look like scrambled phases and intensity distribution, the information is present in every portion of the captured field. Therefore, any small portion of the field containing sufficient speckles is adequate for beam recognition. This opens the possibility of non-line-of-sight optical communication where the detectors can be placed at any off-axis position which is not possible with direct beam detection methods[9]. Also, these methods are robust to noises such as atmospheric turbulence noise and Gaussian white noise. Astigmatic transformation in the speckle field allows the speckle-based model to recognize the intensity degenerate OAM modes, allowing us to utilize the full OAM spectrum[7].

## 2. Generation of Structured Light Speckles:

### 2.1 Simulation

The family of $LG_{p,l}$ and $HG_{m,n}$ modes are the solutions of free space paraxial wave equations in cylindrical and cartesian coordinate systems respectively. Mathematically they are expressed as,

$$LG_{p,l}(\rho,\phi,z) = A_0 \left(\frac{w_0}{w(z)}\right)\left(\frac{\rho}{w(z)}\right)^{|l|} L_p^{|l|}\left(\frac{2\rho^2}{w^2(z)}\right) \times \exp\left(-\frac{\rho^2}{w(z)^2}\right) \exp\left[-i\left(kz + k\frac{\rho^2}{2R(z)}\right)\right] \\ \times \exp(il\phi) \exp[i(|l| + 2p + 1)\Phi(z)] \quad (1)$$

where, $A_0$ is amplitude, $R(z)$ is the wavefront curvature of the beam, $w(z)$ is the effective width of the beam, $w_0$ is the beam waist at $z = 0$, $\Phi(z)$ is the Gouy phase shift, $p$ is a non-negative integer, $l$ is an integer and $L_p^{|l|}$ is an associated Laguerre function of order $p$ and $l$.

$$HG_{m,n}(x,y,z) = A_0\left(\frac{w_0}{w(z)}\right) H_m\left(\frac{\sqrt{2}x}{w(z)}\right) H_n\left(\frac{\sqrt{2}y}{w(z)}\right) \times exp\left[-i\left(kz - k\frac{x^2+y^2}{2R(z)}\right)\right] \\ \times exp[i(m + n + 1)\Phi(z)] \quad (2)$$

Here, $m$ and $n$ are positive integers and represent the number of nodes along $x$ and $y$ axis, $H_m$ and $H_n$ are Hermite polynomials of order $m$ and $n$, respectively.

The far-field speckle pattern of the structured light beams can be generated by passing them through the diffuser – equivalent to multiplying the structured light field with a random phase function that takes the value between 0 and $2\pi$, and then computing the Fourier transform,

$$U^{sc}(r) = \mathcal{F}\left\{U(r) \times exp\left(i\phi_{R_i}(r)\right)\right\}; U^{sc} \in \{LG_{p,l}, HG_{m,n}\} \quad (3)$$

### 2.2 Experiment

Fig. 1a shows the experimental setup for generating the structured light speckle field. The desired LG or HG modes are generated by illuminating the Gaussian laser beam on a phase-only spatial light modulator (SLM) displaying a fork hologram of the desired mode. As the SLM is polarization-sensitive, the polarizer is used to set calibrate the polarization axis of the laser beam. An aperture is used to isolate the desired mode from higher modes appearing at higher diffraction orders. The isolated beam is passed through the rotating diffuser to generate the speckle patterns. The far-field speckle patterns are captured at the back focal plane of the lens (spherical or cylindrical), where the speckle field is fully developed as shown in Fig. 1b. The rotation speed of the diffuser is kept less than the camera exposure time to capture the blur-free and high-contrast speckles.

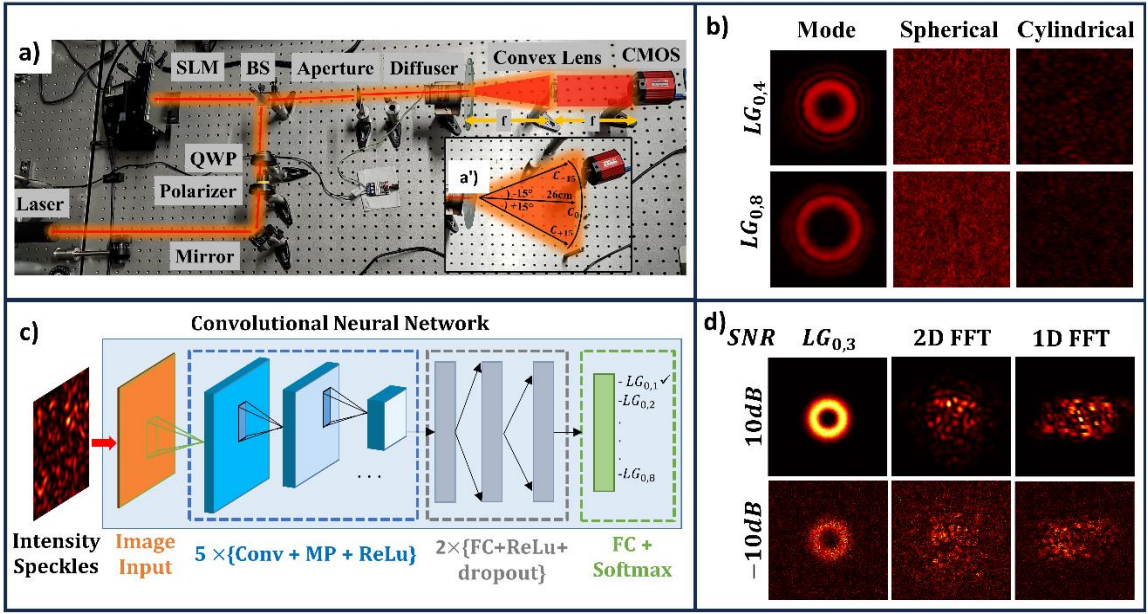

**Fig. 1.** a) Experimental set-up for capturing structured light speckles. Inset: speckle field captured for off-axis recognition. b) Experimentally captured modes and speckle patterns using spherical and cylindrical lens. c) AlexNet architecture. d) Simulated Noisy LG modes simulated with 1D and 2D FFT speckle patterns.

## 3. Speckle-based Structured Light Recognition:

### 3.1 On-axis Structured Light Recognition

A spherical lens has been used to capture the far field of the generated speckle field; analogously these can be generated by taking a two-dimensional Fourier transform ($\mathcal{F}_{2D}\{\ \}$) in Eq. (3). A 1000 speckle pattern images have been generated for each $LG_{p,l}$ ($p = 0$ and $l = 1$ to 8) and 36 $HG_{m,n}$ ($m, n = 0$ to 5) modes, in total 8000 and 36000 speckle pattern images for $LG_{p,l}$ and $HG_{m,n}$ beams respectively. The data has been fed to Alexnet, for training a classification model. Alexnet is a pre-trained CNN known for its moderate computational load and high accuracy. It has a set of Convolutional, ReLu, and pooling layers which extract different features in the data as the images propagate through these layers. The architecture of the Alexnet is shown in Fig.1c. We have obtained an accuracy of $> 99\%$ for simulation data of both $LG_{p,l}$ and $HG_{m,n}$ modes, whereas the model can classify experimental images with an accuracy of 96%.

### 3.2. Astigmatic Transformed Speckle-based Recognition

For a given value of $p$, Eq. (1) gives the identical intensity profiles for modes $LG_{p,l}$ and $LG_{p,-l}$. When 8 pairs of such intensity degenerate modes, $LG_{p,\pm l}$ ($p = 0$ and $l = \pm 1$ to $\pm 8$) are used for training, the model fails to recognize these mutually conjugate modes. This can be overcome by introducing the astigmatic transformation in the speckle field by breaking the symmetry in degenerate speckle distribution. In experiments, it can be achieved by replacing the spherical lens with a cylindrical lens in Fig.1; equivalent to taking a one-dimensional Fourier Transform ($\mathcal{F}_{1D}\{\ \}$) in Eq. (3).

### 3.3 Off-axis Structured Light Recognition

Exploiting the diffractive nature of the speckle allows us to recognize the structured light mode in the off-axis direction too. This allows up to capture speckles at the off-axis position as speckles spread out after scattering from the diffuser. Fig. 1(a') shows the speckles captured at the on-axis position ($C_0$) and at 15° on either side of the optical axis ($C_{\pm 15}$). In this case, we first extract the features from speckle images of various modes using a wavelet scattering network (WSN). The corresponding features given by scattering coefficients are fed to 1D CNN for the training classification model of LG and HG modes [9,10]. The same Alexnet can also be trained for these data but extracting the feature beforehand the model training accelerates the training process and improves accuracy especially when the dataset is small. We achieved an accuracy of 96% and 93% for an on-axis ($C_0$) and off-axis ($C_{\pm 15}$) position for both LG and HG beams. Moreover, it can also classify the intensity degenerate modes with an accuracy of 96% along the optical axis. This shows the broadcasting ability of structured light in short-distance non-line-of-sight communication. The achieved results for all three cases: spherical lens, cylindrical lens, and no lens are given in Table 1.

**Table 1.** Classification accuracies of models trained for recognition of HG and LG modes.

| Structured Light Beams | On-axis Recognition (simulation) | | Off-axis Recognition (experimental) | | |
|---|---|---|---|---|---|
| | Spherical lens | Cylindrical lens | $C_{-15}$ | $C_0$ | $C_{15}$ |
| $HG_{m,n}$ | 99% | 99% | 92% | 96% | 93% |
| $LG_{p,+l}$ | 99% | 99% | 95% | 99% | 93% |
| $LG_{p,\pm l}$ | 50% | 99% | 79% | 96% | 76% |

### 3.4 Robustness Against the Noise

Light beams are prone to disturbances while propagating through any noisy medium. Various kinds of disturbances called noises, can be quantified depending on the propagating medium and other physical conditions. For example, in free-space communication links, light beams suffer through an atmospheric turbulence effect that reduces the accuracy and fidelity of receiving systems. Another type of noise, Gaussian White Noise (GWN), is an inevitable background noise that is present in every system. It is found that the speckle-based recognition models are prone to both kinds of noises[5,7], as shown in Fig1d. The models can classify both LG and HG modes with an accuracy of 98% when trained with modes distorted by passing them through a turbulence phase screen synthesized with $C_n^2 = 1 \times 10^{-14}$ (Refractive index structure parameter). These networks are also robust to Gaussian White Noise as they return classification accuracy of 98% when the SNR of input modes is -10dB.

## 4. Summary and Outlook

In summary, the speckle-learned CNNs provide an alignment and noise-free approach for recognizing structured light modes. The CNNs learn the information present in the speckle distribution which is unique for each mode. The intensity degenerate modes can also be classified when the astigmatic transformation is introduced in the field by passing them through a cylindrical lens. As the speckle field spreads after scattering, it allows us to detect the modes at any off-axis position making the technique fully alignment-free and demonstrating the broadcasting capability of structured light for non-line-of-sight optical communication links. The proposed image-based speckle classification is capable to classify speckle patterns from Fresnel to Fraunhoffer region with high classification accuracy [12]. Moreover, this speckle-based method is not limited to coherent light, it also works with partially coherent light sources[11]. Furthermore, the demonstrated scheme can be easily extended to the near-infrared and microwave-to-radio spectra.

It is important to note that this speckle-learned method is distinct from noise removal techniques, even though they employ similar deep neural layers and activation functions. In this method, the CNN layers are trained to identify underlying features of structured light. Since noise typically lacks a specific pattern, it is automatically disregarded during training without any loss of generality.

**Funding:** VK acknowledges the SERB grant (SRG/2021/001375).

**Acknowledgment:** We acknowledge Prof. Nirmal K. V, and Prof. R.P. Singh for experimental data.